\documentclass[aps,pra,reprint,superscriptaddress,amsfonts,amssymb,amsmath]{revtex4-1}

\usepackage{graphicx}% Include figure files
\usepackage{dcolumn}% Align table columns on decimal point
\usepackage{bm}% bold math
\usepackage[mathlines]{lineno}% Enable numbering of text and display math
\usepackage{siunitx}
\usepackage{booktabs}
\usepackage{color}
\usepackage{mathrsfs}

\begin{document}

\title{Charge-carrier behavior in Ba-, Sr- and Yb-filled CoSb\textsubscript{3}: NMR and transport studies}

\author{Yefan Tian}
\affiliation{Department of Physics and Astronomy, Texas A\&M University, College Station, TX 77843, USA}
\author{Ali A. Sirusi}
 \altaffiliation[Now at ]{Doty Scientific Inc., Columbia, SC.}
\affiliation{Department of Physics and Astronomy, Texas A\&M University, College Station, TX 77843, USA}
\author{Sedat Ballikaya}
\affiliation{Department of Electrical and Electronics Engineering, Istanbul University, Avcilar, Istanbul 34320, Turkey}
\author{Nader Ghassemi}
\affiliation{Department of Physics and Astronomy, Texas A\&M University, College Station, TX 77843, USA}
\author{Ctirad Uher}
\affiliation{Department of Physics, University of Michigan, Ann Arbor, MI 48109, USA}
\author{Joseph H. Ross, Jr.}
\email{jhross@tamu.edu}
\affiliation{Department of Physics and Astronomy, Texas A\&M University, College Station, TX 77843, USA}
\affiliation{Department of Materials Science and Engineering, Texas A\&M University, College Station, TX 77843, USA}

\date{\today}

\begin{abstract}
We report $^{59}$Co NMR and transport measurements on $n$-type filled skutterudites Ba\textsubscript{x}Yb\textsubscript{y}Co\textsubscript{4}Sb\textsubscript{12} and $A$\textsubscript{x}Co\textsubscript{4}Sb\textsubscript{12} ($A$= Ba, Sr), promising thermoelectric materials. The results demonstrate consistently that a shallow defect level near the conduction band minimum dominates the electronic behavior, in contrast to the behavior of unfilled CoSb$_3$. To analyze the results, we modeled the defect as having a single peak in the density of states, occupied at low temperatures due to donated charges from filler atoms. We fitted the NMR shifts and spin-lattice relaxation rates allowing for arbitrary carrier densities and degeneracies. The results provide a consistent picture for the Hall data, explaining the temperature dependence of the carrier concentration. Furthermore, without adjusting model parameters, we calculated Seebeck coefficient curves, which also provide good consistency. In agreement with recently reported computational results, it appears that composite native defects induced by the presence of filler atoms can explain this behavior. These results provide a better understanding of the balance of charge carriers, of crucial importance for designing improved thermoelectric materials.
\end{abstract}

\maketitle

\section{Introduction}

Filled skutterudites $M\textsubscript{z}T_4X_{12}$, where $M$ is a guest atom such as Ba or Yb, $T$ represents a transition metal (Co, Rh, Ir or Fe), and $X$ is a pnicogen or chalcogenide, have gained considerable attention due to their outstanding thermoelectric performance, as well as superconductivity, magnetic ordering, unusual metal-insulator transitions, and heavy fermion behavior \cite{ballikaya2011thermoelectric,meisner1981superconductivity,leithe2004weak,sekine1997metal,torikachvili1987low,luo2015large}. Filled skutterudites obey the phonon-glass electron-crystal concept explaining the significant reduction in thermal conductivity \cite{rowe1995new}. This behavior was first experimentally observed by Morelli \textit{et al.} \cite{morelli1995low}, as a lattice thermal conductivity decreases due to loosely bonded Ce guests. Although the guest atoms act as dopants in addition to inducing low thermal conductivity, they can also modify the electronic behavior, for example, 4$f$ states can cause flat bands and lower the carrier mobility. 

Recently, filled CoSb$_3$-based materials have been studied intensely due to their excellent thermoelectric response \cite{lee2018simple,benyahia2018high,chen2017enhanced,rogl2015doped}. Therefore, the native electronic defects in these skutterudites have been particularly interesting because of their crucial importance for improving thermoelectric efficiency \cite{li2017defect,realyvazquez2017effect,li2016p,xi2015complex,park2011atomic,park2010ab}.

In this work, we utilized NMR, a powerful technique to detect the electronic properties of semiconductors and thermoelectric materials. We describe results for Ba\textsubscript{x}Yb\textsubscript{y}Co\textsubscript{4}Sb\textsubscript{12} and $A$\textsubscript{x}Co\textsubscript{4}Sb\textsubscript{12} ($A$ = Ba, Sr), combined with transport measurements. The results provide new information about the electronic behavior and the importance of defects, very close in energy to the conduction band edge in these $n$-type filled skutterudites.

\section{Sample Preparation and Experimental Methods}

\begin{figure}
\includegraphics[width=0.9\columnwidth]{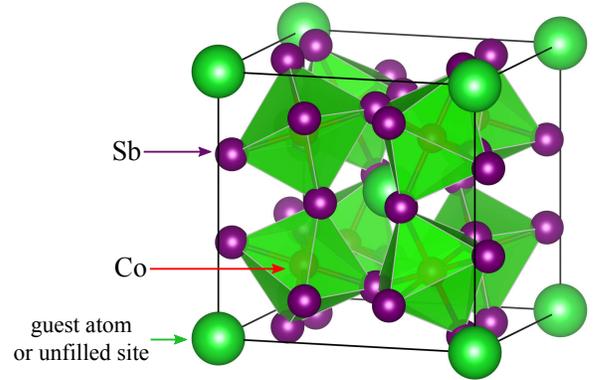}
\caption{\label{skutterudite_structure} Crystal structure of filled skutterudites, $M\textsubscript{z}T_4X_{12}$, where $M$ is a guest atom $(0\leqslant z \leqslant 1)$.}
\end{figure}

Skutterudites (space group $Im3$, shown in Fig.~\ref{skutterudite_structure}, visualized with VESTA \cite{momma2011vesta})
of nominal composition Ba$_{0.1}$Yb$_{0.2}$Co${_4}$Sb$_{12}$, Ba$_{0.2}$Co${_4}$Sb$_{12}$, and Sr$_{0.2}$Co${_4}$Sb$_{12}$ were prepared by a melting-annealing-spark plasma sintering method. The high-purity elements were weighed and mixed in the stoichiometric ratio, loaded into graphite-coated quartz tubes, and sealed in vacuum. These were placed in a furnace and heated to 1373 K at the rate of 1 K\,min$^{-1}$ and held for 10 h to ensure thorough mixing of the constituents. The temperature then decreased to 1013 K at the rate of 4 K\,min$^{-1}$ and held for 10 days. The furnace was then turned off and cooled to room temperature. The resulting ingots were ground to fine powders and pressed by a cold press into pellets, which were then loaded into quartz tubes again, sealed in vacuum, heated to 1013 K at the rate of 4 K\,min$^{-1}$, and held for 2 weeks to form a pure skutterudite phase. The furnace was then turned off and cooled to room temperature. The ingots were ground again to a fine powder and loaded into a graphite die for spark plasma sintering. The final sintering (for densification) was carried out at 903 K for 15 min. Ba$_{0.1}$Yb$_{0.2}$Co${_4}$Sb$_{12}$ is the same sample prepared as reported in Ref.~\cite{ballikaya2011thermoelectric} with $ZT$ approaching 1 at 800 K. 

Electron probe microanalysis (EPMA) measurements including wavelength dispersive spectroscopy were performed to measure the compositions. Results indicate uniform skutterudite phases with very small composition variations. The actual compositions are listed in Table~\ref{table1}.
\begin{table}
\caption{\label{table1}Nominal composition, actual composition (from EPMA analysis), and $n$-type carrier concentration $n_H$ derived from Hall measurements at room temperature.}
{\footnotesize
\begin{ruledtabular}
\begin{tabular}{ccc}
\addlinespace[1ex]Nominal&Actual&$n_H$ ($10^{20}$ cm$^{-3}$)\\[1ex] \hline
\addlinespace[1ex]Ba$_{0.1}$Yb$_{0.2}$Co$_4$Sb$_{12}$&Ba$_{0.07}$Yb$_{0.067}$Co$_4$Sb$_{11.94}$&2.29\\[0.5ex]
\addlinespace[1ex]Ba$_{0.2}$Co$_4$Sb$_{12}$&Ba$_{0.036}$Co$_4$Sb$_{11.77}$&1.80\\[0.5ex]
\addlinespace[1ex]Sr$_{0.2}$Co$_4$Sb$_{12}$&Sr$_{0.041}$Co$_4$Sb$_{11.87}$&1.00\\[0.5ex]
\end{tabular} 
\end{ruledtabular}}
\end{table}
Compared to the nominal compositions, filling fractions are smaller than the starting compositions, which is typical for CoSb$_3$-based skutterudites \cite{mi2011multitemperature}, with remaining filling elements expected to form small oxide particles. In this work, we denote Ba$_{0.1}$Yb$_{0.2}$Co$_4$Sb$_{12}$ as sample Ba(0.1)Yb(0.2), Ba$_{0.2}$Co$_4$Sb$_{12}$ as sample Ba(0.2) and Sr$_{0.2}$Co$_4$Sb$_{12}$ as sample Sr(0.2). Based on the filler atom densities, and assuming ion charges Yb$^{3+}$, Ba$^{2+}$ and Sr$^{2+}$, the measured compositions correspond to $n = 4.6$, $1.0$, and $1.1 \times 10^{20}$ cm$^{-3}$, for Ba(0.1)Yb(0.2), Ba(0.2) and Sr(0.2), respectively, if the ionized charges are donated to the conduction band. However, these can be reduced by native defects, mostly due to departure from CoSb$_3$ stoichiometry. The measured compositions indicate Co excess and Sb deficit typical for these materials, and comparable to the concentrations of filler ions, although note that these composition differences are on the same order as the absolute accuracy of the microprobe. Magnetic measurements were performed using a Quantum Design superconducting quantum interference device
magnetometer. The magnetic results are shown in the Supplemental Material \cite{sm,morelli1995low-t,yang2000iron,chen2016structural,li1994calorimetric,garcia2008direct}, indicating that the samples are non-magnetic with a dilute paramagnetic response attributed to native defects, as well as Yb$^{3+}$ moments in the case of sample Ba(0.1)Yb(0.2).

NMR measurements were carried out by applying a custom-built system at magnetic field 9 T from 4 K to 450 K. $^{59}$Co (nuclear spin $I=7/2$) NMR spectra were obtained using a spin echo sequence with aqueous K$_3$[Co(CN)$_6$] as shift reference. The spin-lattice relaxation times at the central transition lines were determined from fitting to a multi-exponential function for inversion recovery. QuadFit \cite{kemp2009quadfit} was used to fit the spectra. High-temperature transport measurements were carried out under dynamic argon flow in the range 300 to 800 K. Carrier concentrations and Seebeck coefficients were measured using a home-made apparatus with a standard four-probe configuration. Room-temperature carrier concentrations of all samples from Hall measurements are also listed in Table~\ref{table1}. As opposed to the $p$-type behavior of pure CoSb$_3$, all samples show $n$-type behavior due to electron donation from filler atoms.

\section{Experimental results}

\subsection{NMR measurements}

\subsubsection{Line shapes}

\begin{figure}
\includegraphics[width=\columnwidth]{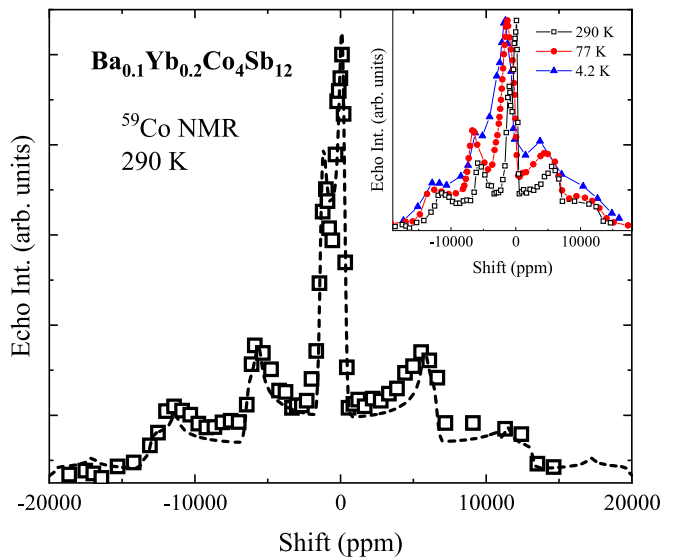}
\caption{\label{LD_CoSb3_lineshape} $^{59}$Co NMR spectrum for sample Ba(0.1)Yb(0.2) at 290 K. The dashed line is a fit for the overall spectrum. The inset shows spectra at 4.2 K, 77 K, and 290 K, normalized to the same peak intensity.}
\end{figure}

\begin{figure}
\includegraphics[width=0.9\columnwidth]{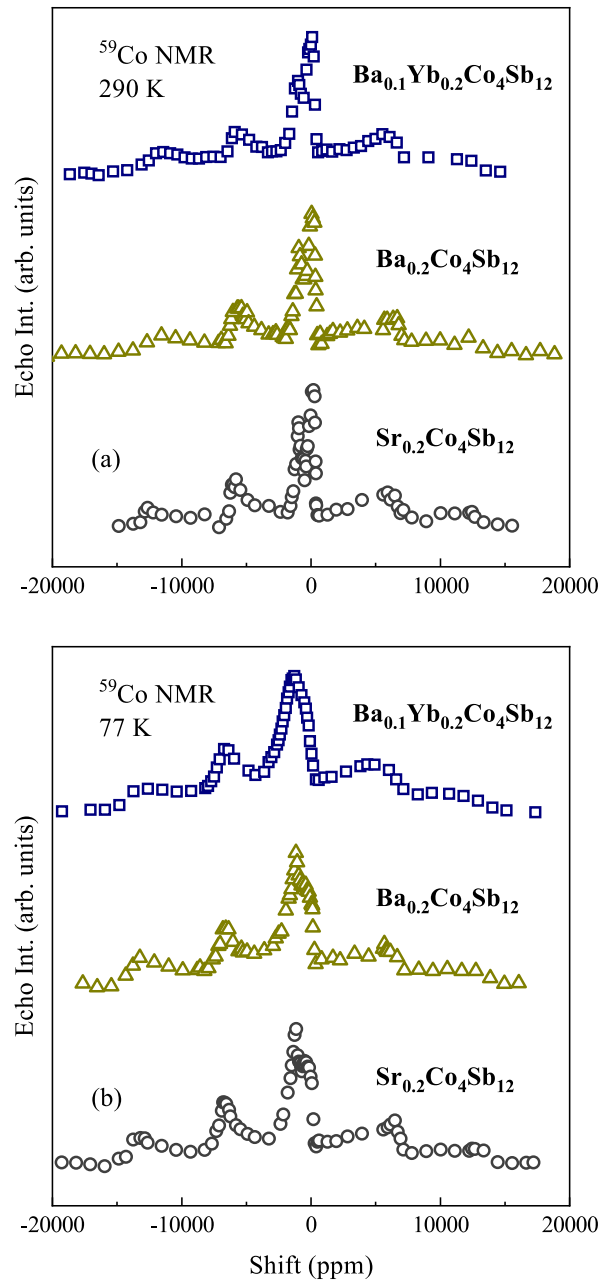}
\caption{\label{LD_CoSb3_all_lineshapes} $^{59}$Co NMR spectra for samples Ba(0.1)Yb(0.2), Ba(0.2) and Sr(0.2) in (a) 290 K and (b) 77 K. The data are offset vertically for clarity.}
\end{figure}

Fig.~\ref{LD_CoSb3_lineshape}
shows the $^{59}$Co NMR spectrum of sample Ba(0.1)Yb(0.2) at 290 K. The powder spectra exhibit the characteristic sequence of edge singularities due to the $\Delta m=1$ nuclear transitions. The quadrupole frequency is defined as $\nu_Q=3eQV_{zz}/[2I(2I-1)]$, where $Q$ is the nuclear quadrupole moment and $V_{zz}$ is the maximum principal value of the electric field gradient (EFG) tensor. The other two principal values $V_{xx}$ and $V_{yy}$ are equal due to the axial Co site symmetry. Compared to CoSb$_3$ with $\nu_Q =1.18$ MHz \cite{lue2007nmr}, the fitted room-temperature spectra (dashed line in Fig.~\ref{LD_CoSb3_lineshape}) have $\nu_Q=1.11\pm0.01$ MHz for all 3 samples. The chemical shift anisotropies for Ba(0.1)Yb(0.2), Ba(0.2) and Sr(0.2) are 1250 ppm, 1200 ppm and 1220 ppm, respectively (given as the span, $\Omega=\delta_{11}-\delta_{33}$). This is in good agreement with the value $K_\text{ax}=-0.039\%$ (corresponding to $\Omega=1170$ ppm) for CoSb\textsubscript{3} reported by Lue \textit{et al.} \cite{lue2009evolution}. The reduction in $\nu_Q$ due to filler atoms is similar to the behavior of Ca\textsubscript{x}Co\textsubscript{4}Sb\textsubscript{12} and La\textsubscript{x}Co\textsubscript{4}Sb\textsubscript{12} \cite{lue2009evolution,lue2008effect}. The inset of Fig.~\ref{LD_CoSb3_lineshape} displays spectra for sample Ba(0.1)Yb(0.2) at 4.2 K, 77 K, and 290 K, demonstrating the shift to lower frequencies along with increasing line width upon cooling.
Fig.~\ref{LD_CoSb3_all_lineshapes} 
shows a comparison of spectra for all three samples at room temperature and 77 K.

Note that in the measured frequency range, $^{121}$Sb signals coexsit with those of $^{59}$Co, however with very large line widths. Based on NQR in La-filled CoSb\textsubscript{3}, a large $\nu_Q$ of 38.8 MHz has been reported \cite{magishi2014nmr}. Similar results for the present materials will lead to a central transition for $^{121}$Sb $\sim$1200 times wider than the $^{59}$Co central transition, or about 10 MHz, and the spectral intensity correspondingly reduced by a very large factor.

\subsubsection{Knight shifts}

\begin{figure}
\includegraphics[width=0.9\columnwidth]{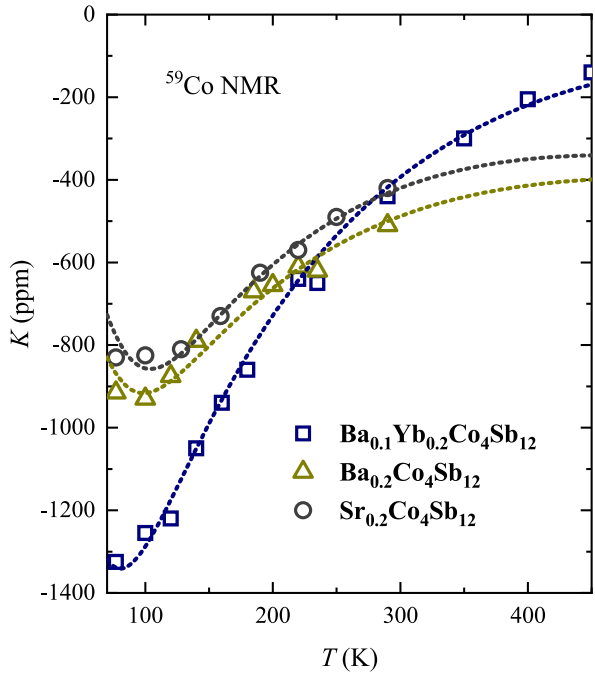}
\caption{\label{LD_CoSb3_shiftvsT} Temperature-dependent $^{59}$Co NMR shift for samples Ba(0.1)Yb(0.2), Ba(0.2) and Sr(0.2). The dotted lines are fits based on the model described in the text.}
\end{figure}

Fig.~\ref{LD_CoSb3_shiftvsT}
shows the temperature-dependent $^{59}$Co NMR isotropic shift obtained for the three samples by calculating the center of mass of the central transition ($-1/2\leftrightarrow+1/2$) portion of the spectrum. The shift is composed of two main contributions: Knight shift and chemical shift. The Knight shift, due to unpaired spins of charge carriers and defects, provides the large temperature dependence for these samples. Thus, for simplicity, in the plot $K$ is used to represent the entire shift, similar to what has been reported for La\textsubscript{x}Co$_4$Sb$_{12}$ \cite{lue2008effect}. It can be seen that the total shift is negative, becoming more positive with temperature increasing. In Sec.~\ref{general}, we will discuss a theoretical model for this temperature dependence in terms of increasing carrier density vs. temperature. Due to the increasing shift and broadening, close to 4 K the central transition and satellite transitions peaks merge for all three samples, which makes it difficult to isolate the shift at very low temperatures.

\subsubsection{Spin-lattice relaxation rates}

\begin{figure}
\includegraphics[width=\columnwidth]{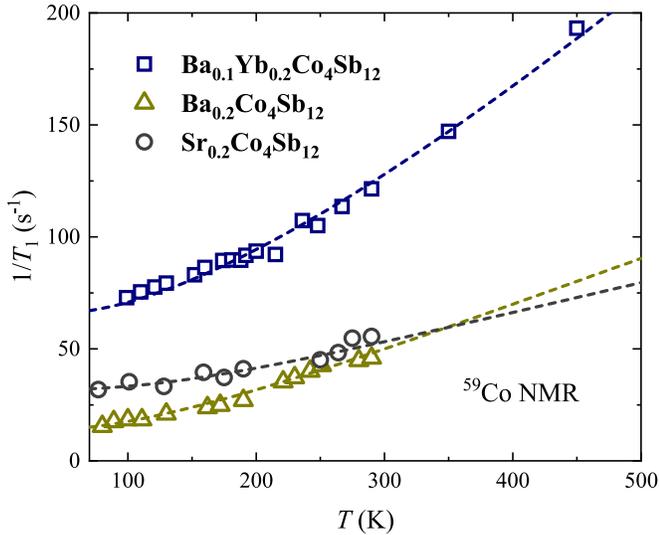}
\caption{\label{LD_CoSb3_T1} $1/T_1$ vs. $T$ for samples Ba(0.1)Yb(0.2), Ba(0.2) and Sr(0.2). The dashed lines are fits as described in the text.}
\end{figure}

The $^{59}$Co spin-lattice relaxation rate was measured using the inversion recovery method, based on the integrated spin echo fast Fourier transform of the $^{59}$Co ($I = 7/2$) lines. We irradiated only the central portion of the spectra corresponding to the peak intensity of the $-1/2\leftrightarrow+1/2$ transitions, well known to give a multi-exponential recovery. Similar to the shift data, this was done only at 77 K and above, due to the merging of the central transition with satellite transitions at low $T$. For the central transition with $I = 7/2$, the recovery of the nuclear magnetization due to spin excitations can be expressed as
\begin{multline} \label{T1}
\frac{M(t)-M(\infty)}{M(\infty)}=-2\alpha(0.012e^{-\frac{t}{T_1}}+0.068e^{-\frac{6t}{T_1}}\\
+0.206e^{-\frac{15t}{T_1}}+0.714e^{-\frac{28t}{T_1}}).
\end{multline}
Here, $\alpha$ is a fractional value derived from the initial conditions, $M(t)$ is the nuclear magnetization at time $t$, and $M(\infty)$ represents the asymptotic signal. Each experimental value was obtained by a fit to Eq.~(\ref{T1}). The resulting $1/T_1$ values are shown in Fig.~\ref{LD_CoSb3_T1}.

\subsection{Transport measurements}

\begin{figure}
\includegraphics[width=\columnwidth]{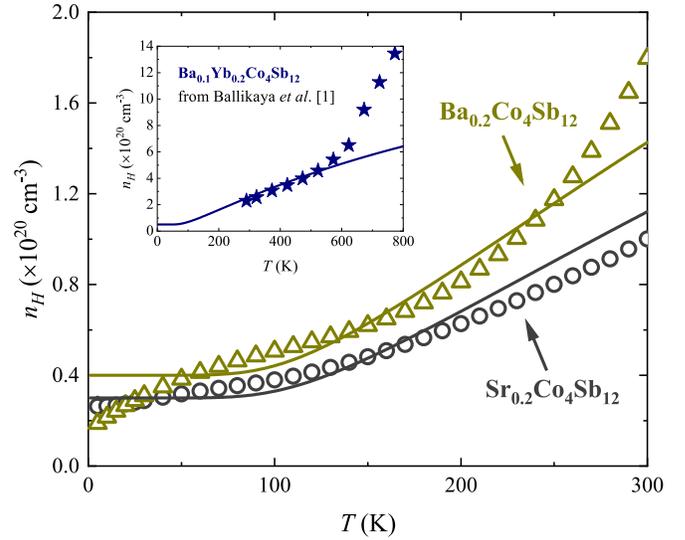}
\caption{\label{LD_CoSb3_carrier} Carrier concentration vs. $T$ from 4 K to 300 K for samples Ba(0.2) and Sr(0.2). Inset shows $n_H$ vs. $T$ for sample Ba(0.1)Yb(0.2) from room temperature to 773 K \cite{ballikaya2011thermoelectric}. Solid curves: conduction-band carrier concentration ($n_\text{CB}$); model described in text.}
\end{figure}

\begin{figure}
\centering
\includegraphics[width=0.75\columnwidth]{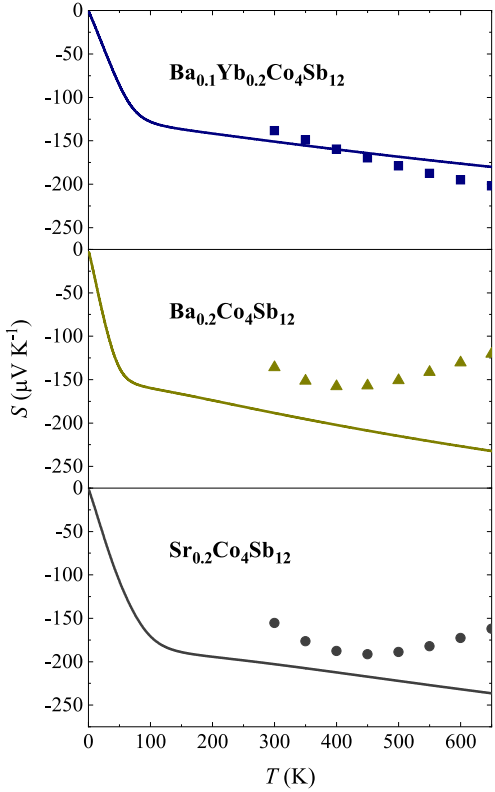}
\caption{\label{LD_CoSb3_seebeck} Seebeck coefficient vs. $T$ from 300 K to 650 K for samples Ba(0.1)Yb(0.2), Ba(0.2) and Sr(0.2). Theoretical curves are shown superposed on each plot.}
\end{figure}

Carrier concentrations ($n_H$) for samples Ba(0.2) and Sr(0.2) obtained by Hall measurements from 4 K to 300 K are plotted in Fig.~\ref{LD_CoSb3_carrier},
with the inset showing data from room temperature to 773 K of sample Ba(0.1)Yb(0.2) extracted from Ref.~\cite{ballikaya2011thermoelectric}.
Fig.~\ref{LD_CoSb3_seebeck}
shows the Seebeck coefficients of all samples from 300 K to 800 K. Both of these sets of results have the expected sign corresponding to $n$-type materials. Although the increase in $n_H$ of Fig.~\ref{LD_CoSb3_carrier} could be attributed to excitation of holes and electrons across a very small band gap on order of 30-50 meV, the Seebeck results generally support non-compensated behavior with a larger band gap consistent with other experimental results \cite{nolas1996raman,caillat1996properties} and calculations \cite{wee2010effects,sofo1998electronic,lefebvre2001bonding}, with negligible hole contribution at temperatures of the NMR measurements. The decrease in $S$ above 400 K observed for the Ba- and Sr-filled samples could possibly indicate the excitation of holes above this temperature \cite{goldsmid1999estimation}, although there are also alternative explanations for this behavior as will be discussed in Sec.~\ref{discussion}.

\section{Theoretical modeling and analysis} \label{general}

To analyze the experimental data, we developed a formalism for the interaction between nuclei and carriers in the conduction band, allowing for arbitrary carrier densities rather than treating the extreme metallic or non-degenerate limit. The model assumes that all carriers reside in the conduction band.

The Knight shift ($K$) reflects the local effective magnetic field at the nuclei due to conduction electrons, given by
\begin{equation} \label{define_knight_shift}
K=\frac{\Delta\nu}{\nu_0}=\frac{H_\text{HF}\chi_s^e}{\mu_B},
\end{equation}
where $H_\text{HF}$ represents a hyperfine field. Note that conduction band edge has been shown to be dominated by Co $d$-states \cite{puyet2006electronic}. If the interaction is Fermi contact, this is defined by $H_\text{HF}=\frac{8\pi}{3}\langle|\varphi_k(0)|^2\rangle_{E_F}\mu_B$, however, since $d$ electrons are dominant here, $H_\text{HF}$ is instead the core polarization hyperfine field \cite{narath1967hyperfine,walstedt2018nmr}. For $^{59}$Co, we used the measured value $-21.7$ T \cite{janak1979calculated}. $\chi_s^e$ is the electron spin susceptibility which can be calculated according to the average carrier concentration per Co atom $N_\text{atom}(\mu)=n(\mu)\times V_\text{atom}$,
\begin{equation} \label{chi_spin}
 \chi_s^e \cong \frac{\mu_B}{2H}(g^*\mu_BH)\frac{\partial N_\text{atom}}{\partial\mu} = \frac{g^*}{2}\frac{\partial N_\text{atom}}{\partial\mu}\mu_B^2,
\end{equation}
in which $V_\text{atom}$ is the average volume per Co atom and $\mu$ is the chemical potential. $n(\mu)$ is defined by
\begin{equation} \label{define_n}
n(\mu)\equiv\int g(E)f(E, \mu)dE,
\end{equation}
where the conduction band density of states is given by
\begin{equation} \label{dos}
g_\text{CB}(E)=
\begin{cases}
\frac{\sqrt{2}Nm^{*^{3/2}}}{\pi^2\hbar^3}\sqrt{E-E_C}, E \geqslant E_C \\
0, E < E_C
\end{cases}
\end{equation}
with $N$ the number of minima and $m^*$ the effective mass in the band edge, assumed to be parabolic. The Fermi function is,
\begin{equation} \label{fermi}
f(E, \mu)=\frac{1}{e^{\frac{E-\mu}{kT}}+1}.
\end{equation}
By substituting Eqs.~(\ref{chi_spin}), (\ref{define_n}), (\ref{dos}) and (\ref{fermi}) into Eq.~(\ref{define_knight_shift}) and letting $x=E-E_C$, we can derive
\begin{equation} \label{knight_shift}
K=K_0+\frac{A_1}{T}\int^\infty_0\frac{\sqrt{x}e^{-\frac{E_C-\mu+x}{kT}}}{(e^{-\frac{E_C-\mu+x}{kT}}+1)^2}dx,
\end{equation}
with $A_1=H_\text{HF}g^*\mu_BNV_\text{atom}m^{*^{3/2}}/\sqrt{2}\pi^2\hbar^3k$ and the constant term $K_0$ representing an additive chemical shift and background Knight shift.

The $T_1$ relaxation process can be understood on the basis of scattering from initial occupied electron states to final unoccupied states. According to Fermi$^\prime$s golden rule, the transition rate from state $i$ to state $f$ is given by \cite{slichter2013principles}
\begin{equation} \label{scattering}
\Gamma_{i\rightarrow f}=\frac{2\pi}{\hbar}|\langle f|V|i \rangle|^2\delta(E_f-E_i),
\end{equation}
where $V=H_\text{HF}\gamma_n\hbar\bf{I\cdot S}$ is the interaction providing the scattering mechanism. As a result, $1/T_1$ can be expressed by \cite{meintjes2005temperature}
\begin{multline} \label{T_1}
\frac{1}{T_1}=\frac{1}{2}\iint\Gamma_{i\rightarrow f}\Big(g_\text{CB}(E_i)f(E_i)\Big)\\
\times\Big(g_\text{CB}(E_f)[1-f(E_f)]\Big)dE_idE_f,
\end{multline}
where $E_i \approx E_f$ represent initial and final states, respectively. By substituting Eqs.~(\ref{dos}), (\ref{fermi}) and (\ref{scattering}) into Eq.~(\ref{T_1}),
\begin{equation} \label{spin_lattice}
\frac{1}{T_1}=\frac{1}{T_{1C}}+A_2\int^\infty_0\frac{xe^{-\frac{E_C-\mu+x}{kT}}}{(e^{-\frac{E_C-\mu+x}{kT}}+1)^2}dx,
\end{equation}
with $A_2=8H_\text{HF}^2\mu_B^2N^2V_\text{atom}^2m^{*^3}\gamma_n^2/\pi^3\hbar^7\gamma_e^2$ and $1/T_{1C}$ representing other contributions to the relaxation rate. Note that in the highly degenerate limit ($\mu-E_C\gg kT$), these results can easily be shown to simplify to $K=\text{const.}$ and $1/T_1\propto T$, as often seen for heavily doped semiconductors \cite{sirusi2017recent}. Here, we consider the more general case, since $\mu-E_C\approx kT$ for much of the range considered here.

The carrier concentration in the conduction band, $n_\text{CB}$, can be derived by substituting Eqs.~(\ref{dos}) and (\ref{fermi}) into Eq.~(\ref{define_n}),
\begin{equation} \label{carrierconc}
n_\text{CB}=\frac{\sqrt{2}Nm^{*^{3/2}}}{\pi^2\hbar^3}\int_{0}^{\infty}\frac{\sqrt{x}}{e^{\frac{E_C-\mu+x}{kT}}+1}dx.
\end{equation}
Also the Seebeck coefficient can be calculated by $S=-\frac{1}{eT}\frac{\mathscr{L}^{(1)}}{\mathscr{L}^{(0)}}$, where $\mathscr{L}^{(\alpha)}_{ij} \equiv e^2 \int \frac{d^3\bf{k}}{4\pi^3}(-\frac{\partial f}{\partial E})\tau v_i v_j (E-\mu)^\alpha$ \cite{ashcroft1976solid}. By substituting Eqs.~(\ref{dos}) and (\ref{fermi}), the Seebeck coefficient is expressed as
\begin{equation} \label{seebeck}
\begin{split}
 S & = -\frac{1}{eT}\frac{\int (df/dE)g(E)\tau(E)E(E-\mu)dE}{\int (df/dE)g(E)\tau(E)EdE} \\
 & = -\frac{1}{eT}\frac{\displaystyle \int^\infty_0 \frac{e^{-\frac{E_C-\mu+x}{kT}}}{(e^{-\frac{E_C-\mu+x}{kT}}+1)^2}\tau(x)x^{3/2}(x-\mu)dx}{\displaystyle \int^\infty_0 \frac{e^{-\frac{E_C-\mu+x}{kT}}}{(e^{-\frac{E_C-\mu+x}{kT}}+1)^2}\tau(x)x^{3/2}dx},
\end{split}
\end{equation}
where the second form assumes explicitly that $g_\text{CB}\propto (E-E_C)^{1/2}$. Typically, $\tau(E)$ is considered proportional to $(E-E_C)^r$ with $-3/2\leqslant r \leqslant 1/2$ depending on the scattering mechanism \cite{mahan1989figure}.

\begin{figure}
\includegraphics[width=0.7\columnwidth]{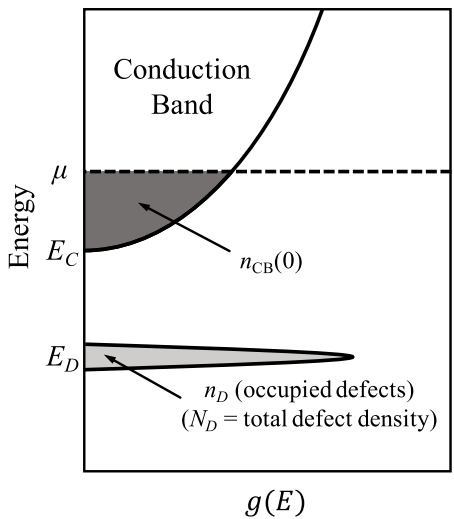}
\caption{\label{LD_CoSb3_DOS} The general model for the total density of states used here. $E_C$, $E_D$ and $\mu$ are the positions of conduction band minimum, shallow defect state and chemical potential, respectively. $n_\text{CB}$ represents the carrier concentration in the parabolic conduction band. For simplicity this is pictured for $T=0$ for which $n_\text{CB}(0)$ is the limiting value. $n_D$ represents the electron concentration in shallow defect states (light gray area), with available level density $N_D$, assumed to be a Dirac delta function.}
\end{figure}

Fig.~\ref{LD_CoSb3_DOS}
shows the simple model for $g(E)$ found to give consistent agreement with the results, with shallow defect states assumed to be represented by a single Dirac delta function due to a superposition of isolated in-gap states at energy $E_D$. This leads to the low-temperature increase in $n_H$, with $\mu$ in the conduction band at $T=0$ due to electrons donated by the filler atoms. The conservation of total charge, $n_D(0)+n_\text{CB}(0)=n_D(T)+n_\text{CB}(T)$, determines the temperature dependence of the chemical potential $\mu$, where $n_D$ represents the electron concentration in shallow defect states (light gray area shown in Fig.~\ref{LD_CoSb3_DOS}). The relationship of $\mu$ and $T$ is thus given by
\begin{equation} \label{conservation}
\begin{split}
N_D+n_\text{CB}(0) & =\int N_D\delta(E-E_D)f(E)dE+n_\text{CB} \\
& = N_D\frac{1}{e^{\frac{E_D-\mu}{kT}+1}}+n_\text{CB},
\end{split}
\end{equation}
where $N_D=n_D(0)$ and $n_D(T)/N_D$ are the concentration of shallow states and the filled fraction at a given $T$, respectively, and $n_\text{CB}(0)$ is the carrier concentration at $T=0$. Then for each temperature, the corresponding chemical potential can be obtained by numerically solving Eq.~(\ref{conservation}).

\begin{table*}
\caption{\label{parameters}Parameters of theoretical fittings for samples Ba(0.1)Yb(0.2), Ba(0.2) and Sr(0.2). $E_D$ is the position of defect state relative to $E_C$, $N_D$ the concentration of defect state, $n_\text{CB} (0)$ the CB carrier concentration at 0 K, $m_\text{eff}=N^{2/3}m^*$ the thermodynamic effective mass, $g^*$ the effective $g$-factor, $K_0$ the addictive shift and $1/T_{1C}$ the spin-lattice relaxation rate at 0 K.}
{\footnotesize
\begin{ruledtabular}
\begin{tabular}{cccccccc}
\addlinespace[1ex]Sample&$E_D$ (meV)&$N_D$ $(10^{20}$ cm$^{-3})$&$n_\text{CB} (0)$ $(10^{20}$ cm$^{-3})$&$m_\text{eff}$ ($m_e$)&$g^*$&$K_0$ (ppm)&$1/T_{1C}$ (s$^{-1}$)\\[1ex] \hline
\addlinespace[1ex]Ba(0.1)Yb(0.2)&$-26$&15&0.5&3.8&$-8$&$-3560$&61 \\[0.5ex]
\addlinespace[1ex]Ba(0.2)&$-35$&6&0.4&3.4&$-8.5$&$-2570$&12 \\[0.5ex]
\addlinespace[1ex]Sr(0.2)&$-38$&4.5&0.3&3.2&$-9$&$-2230$&30 \\[0.5ex]
\end{tabular}
\end{ruledtabular}}
\end{table*}

Figs.~\ref{LD_CoSb3_shiftvsT}-\ref{LD_CoSb3_seebeck} show fitted theoretical curves based on this model. Since the numerical solution of several integral equations is required, we cannot least-squares fit all parameters at once. However, we find that the carrier concentration is much more sensitive to effective mass than the other quantities, so the thermodynamic effective mass, $m_\text{eff}=N^{2/3}m^*$ and $n_\text{CB}(0)$ were fitted to the carrier concentration with values shown in Table~\ref{parameters}. Both for Ba(0.2) and Sr(0.2), $n_H$ exhibits a decrease below about 50 K, apparently a trend toward localization at the temperatures, so we fitted the data above this temperature. The results for $m_\text{eff}$ are in close agreement with each other, as might expected for rigid-band filling of states at the CB edge. Then by fitting $K$ and $1/T_1$ together, $E_D$, $N_D$ and, $g^*$ were optimized, giving the results also listed in Table~\ref{parameters}. Values of $g^*$ are between $-8$ and $-9$, comparable to that of the holes, with smaller $m^*$, having $g^*=-10.1$ as reported by Arushanov \textit{et al.} \cite{arushanov2000shubnikov}.

While the gradual increase of Knight shift and $1/T_1$ vs. temperature can be understood in terms of an increasing number of carriers excited into CB, the terms $K_0$ and $1/T_{1C}$ can be interpreted as due to the susceptibility of electrons in the localized levels. To the extent that Coulomb interactions allow unpaired spins within these states, a Curie-type susceptibility contribution is expected at low temperatures, the presence of which is confirmed by the large negative NMR shifts at helium temperatures in all three samples. These shifts are responsible for the overlapping of transitions at low temperatures, which prevented separation of the central transition shifts in this limit as discussed above. Note that the mechanism involves contact with Co $d$-states through the negative $H_\text{HF}$ for Co, as opposed to dipole coupling for dilute local moments which provides an additional broadening mechanism for Yb$^{3+}$ moments in sample Ba(0.1)Yb(0.2), but no net shift contribution \cite{tian2018native}. Other contributions to the shifts include differences in chemical shift, and above 77 K we find that an added constant term ($K_0$) can best fit the temperature dependence of these shifts.

The localized-electron contribution to $1/T_1$ can be modeled directly in terms of the dynamical susceptibility of such localized spins. From general considerations, it is often found \cite{warren1971nuclear,paalanen1985spins} that $1/T_1 \propto k_B T \chi_0 \tau$, where $\chi_0$ is the DC susceptibility and $\tau$ is an electron spin lifetime. For sufficiently concentrated localized spins, $\tau$ can approach a constant due to spin diffusion, even for carriers which do not contribute to the electrical conductivity, and with $\chi \propto 1/T$, this gives a constant contribution to $1/T_1$. Similar results are obtained for Si:P near its metal-insulator transition \cite{meintjes2005temperature,paalanen1985spins}. The term ($1/T_{1C}$) is thus expected to be due to such a contribution, and it seems reasonable that sample Ba(0.1)Yb(0.2), for which we obtained the largest density of localized states ($N_D$), this contribution to $1/T_1$ is found to be the largest.

The resulting theoretical transport curves for all three samples are shown in Figs.~\ref{LD_CoSb3_carrier} and \ref{LD_CoSb3_seebeck}. With the chemical potentials solved by Eq.~(\ref{conservation}) plugged into Eq.~({\ref{carrierconc}), the theoretical curves describe the temperature-dependent $n_H$ quite reasonably. The deviation of the theoretical $n_H$ from the experimental data above 600 K for sample Ba(0.1)Yb(0.2) is likely due to carriers excited to a second band in higher temperature with a corresponding increase in effective mass \cite{tang2015convergence}. With no adjustable parameters, the theoretical curves for Seebeck coefficient were drawn directly from Eq.~(\ref{seebeck}) as shown in Fig.~\ref{LD_CoSb3_seebeck}. For these plots, $\tau(E)$ was taken to be proportional to $(E-E_C)^r=(E-E_C)^{-1/2}$ due to the acoustic phonon deformation potential mechanism. This mechanism is shown to provide a good agreement for materials with complex structures and multi-valley Fermi surfaces \cite{wang2013material}, although there are indications that in some substituted skutterudites the mechanism and exponent $r$ may change vs. $T$ \cite{sun2015large}; this would have the effect of scaling the resulting $S(T)$ curves vertically. A distribution of defect energies $(E_D)$ would also explain the softer turn-on apparent in the $S(T)$ data, however, our simple model successfully predicts both the sub-linear temperature dependence and the approximate magnitudes of $S(T)$, without adjustment of the parameters.

\section{Discussion} \label{discussion}

The model of Fig.~\ref{LD_CoSb3_DOS} provides a consistent picture of both the transport and NMR results and thus indicates the importance of states near the conduction band edge in filled CoSb$_3$. This differs from unfilled CoSb$_3$, for which native deep acceptor states are believed to dominate the behavior \cite{park2010ab,li2017defect}, although recent experimental evidence also indicates $n$-type behavior for the case of large Sb deficit \cite{realyvazquez2017effect}.

In our model, the valence band is completely filled with negligible hole density over the measured temperature range. Thus we do not probe the band gap from VBM to CBM directly. However, from our results it appears that previous results showing evidence for excitation across a gap of order tens of milli-electron volts \cite{nagao2000electron,arushanov2000shubnikov,lue2007nmr} are likely also dominated by defect levels close to the CB, while the relatively larger band gaps obtained by other techniques \cite{sharp1995thermoelectric,mandrus1995electronic,nolas1996raman,caillat1996properties,kajikawa2014effects} are consistent with what we propose. Computational results based on DFT and more advanced techniques generally indicate a band gap in the range 0.2 to 0.6 eV for CoSb$_3$ \cite{tang2015convergence,park2010ab,khan2015electronic} with relatively small changes due to filler atom densities comparable to those in our samples \cite{wee2010effects,hu2017effects}. Our results thus demonstrate that a larger gap combined with the presence of additional donor states can explain previous inconsistencies in the reported band gap. Note that while the calculated chemical potential positions in this model dip toward the CB edge as the temperature rises, they remain far above the VB edge, such that our assumption of negligible hole density remains valid. With the VB effective mass ratio reported to be $m_h^*/m_e^*=0.24$ \cite{lue2007nmr}, we obtained hole densities by direct integration (e.g. similar to Eq.~(\ref{carrierconc})) for the Sr(0.2) sample, which in the numerical results has the smallest chemical potential. At 300 K, for a band gap at the low end of the range quoted above (0.2 eV) we obtain a hole density $1\times10^{16}$ cm$^{-3}$, and orders of magnitude smaller as the gap increases. Thus over the range of expected behavior the VB has a negligible contribution to the transport and NMR behavior.

Regarding the origin of the defect states shown to sit near the conduction band edge, computational results give a possible explanation based on the presence of composite defects. While off-stoichiometric CoSb$_3$ is usually $p$-type because of acceptor-like defects, Co interstitial pairs are also proposed as $n$-type \cite{park2010ab} or $p$-type \cite{li2017defect} defects. These pairs are believed to form only at temperatures below that of typical processing conditions, however Hu \textit{et al.} recently indicated that La filling combined with Sb di-vacancies can form shallow defect states near the conduction band minimum \cite{hu2017effects}. By analogy with this result, it seems likely that the defects observed here are associated with composite defects induced by the filler atoms (Ba, Sr, and Yb). The fitted values in Table~\ref{parameters} bear this out; donor charges $N_D$ approximately three times larger than the expected filler atom charges point to such composite defects making up the donor states rather than the charges associated with the filler atoms themselves. The difference is comparable to the density of Co excess/Sb deficiency, and thus it appears that the filler atoms tie up these native defects, producing the donor states observed here.

The effective masses for all three samples are quite close to each other and in good agreement with the predicted $m_\text{eff}\approx3.4\,m_e$ from modeling and experimental results for $n=1\times10^{20}$ cm$^{-3}$ as reported by Caillat \textit{et al.} \cite{caillat1996properties}. In Ref.~\cite{tang2015convergence}, a slightly smaller $m_\text{eff}\approx2.8\,m_e$ was obtained for $n=2\times10^{20}$ cm$^{-3}$, but note that this was derived from Seebeck coefficient results assuming a degenerate limit. However, we find that all samples begin to deviate from this limit, along with a non-constant $n_H$, above room temperature.

As one of the most promising thermoelectric systems, a clear picture of the overall electronic structure of CoSb$_3$-based materials can give a better understanding of the transport results, which will directly help to design high-$ZT$ thermoelectric materials. Also, since carrier donation from filler atoms is needed to optimize thermoelectric performance, a good understanding of defect states and impurity bands will be significant for thermoelectric device design. In addition, as we have shown, NMR can be a very effective tool for such analysis.

\section{Conclusions}

NMR and transport results of filled skutterudites Ba\textsubscript{x}Yb\textsubscript{y}Co\textsubscript{4}Sb\textsubscript{12} and $A$\textsubscript{x}Co\textsubscript{4}Sb\textsubscript{12} ($A$ = Ba, Sr) demonstrate the existence of a shallow defect level below conduction band minimum. To fit the experimental results, a simple but effective theoretical model was established by assuming the defect states to be represented by a single narrow peak in the density of states. The NMR and transport results were analyzed in a very general way allowing the Hall effect as well as Knight shift and $T_1$ results to be fitted numerically as the carriers slowly changed from metallic to non-degenerate situation. These fits yielded an effective mass in good agreement with predicted values and indicated that the gradual changes in Hall coefficient observed at low temperatures in filled CoSb$_3$ are associated with a defect state positioned close to the conduction band minimum. In addition, Seebeck coefficient data were also treated within the same general model and found to agree with parameters derived from the other measurements.

\begin{acknowledgments}
This work was supported by the Robert A. Welch Foundation, Grant No. A-1526. Sedat Ballikaya kindly acknowledges financial support by Scientific and Technological Research Council of Turkey (TUBITAK) with project number 115F510 and Scientific Research Projects Coordination Unit of Istanbul University with project number of 21809 and 32641. Synthesis of skutterudites samples at the University of Michigan was supported as part of the Center for Solar and Thermal Energy Conversion, an Energy Frontier Research Center funded by the U.S. Department of Energy, Office of Basic Energy Sciences under Award DE-SC-0000957.
\end{acknowledgments}

\end{document}